\documentclass[a4paper,11pt]{article}
\usepackage{pos}
\usepackage{bbm,color}
\usepackage{mathtools}
\usepackage{graphicx}
\usepackage{epsfig}
\usepackage{euscript}
\usepackage{xcolor}
\usepackage{fancybox}
\usepackage{hyperref}
\usepackage{enumerate}
\usepackage{widetext}
\usepackage{subfigure}
\definecolor{brightturquoise}{rgb}{0.03, 0.91, 0.87}
\definecolor{awesome}{rgb}{1.0, 0.13, 0.32}
\definecolor{armygreen}{rgb}{0.29, 0.33, 0.13}
\definecolor{aqua}{rgb}{0.0, 1.0,1.0}
\definecolor{maroon(html/css)}{rgb}{0.5, 0.0,0.0}
\definecolor{pinegreen}{rgb}{0.0, 0.47,0.44}
\definecolor{red-brown}{rgb}{0.65, 0.16,0.16}

\def\rQCED{{\rm QCED}}

\newcommand{\KK}{${\cal KK}$}

\title{Role of IR-Improvement in LHC/FCC Physics}

\author*[a]{B.F.L. Ward}
\author[a]{B. Shakerin}
\author[b]{S. A. Yost}
\affiliation[a]{Baylor University,\\
   Waco, TX, USA}
\affiliation[b]{The Citadel,\\
Charleston, SC, USA}

\emailAdd{bfl\_ward@baylor.edu}
\emailAdd{bahram.shakerin@gmail.com} 
\emailAdd{yosts1@citadel.edu} 

\abstract{One may use amplitude-based resummation in QED $\otimes$ QCD to achieve IR-improvement of unintegrable singularities in the infrared regime to arbitrary precision in principle. We illustrate such improvement in specific examples in precision LHC/FCC physics.}
\centerline{BU-HEPP-20-05, Dec., 2020}
\FullConference{%
  40th International Conference on High Energy physics - ICHEP2020\\
  July 28 - August 6, 2020\\
  Prague, Czech Republic (virtual meeting)
}

\begin{document}
\maketitle

\baselineskip=11pt
    In the context of QCD, in addition to the original discussion \footnote{F. Berends, private communication, at 1988 ICHEP Conference Dinner, Munich, Germany.} of whether the naive Jackson-Scharre~\cite{js} or the exact YFS-style~\cite{yfs} resummation is more accurate for a given level of exactness, we have the issue of a hard cut-off for the IR versus resummed IR integrability. Somewhat unexpectedly, in the current era of QCD with precision tags $\lesssim 1.0\%$ with accompanying EW precision tags at the per mille level for processes such as single heavy gauge boson production at the LHC, the ``new'' issue of ISR radiation from quarks has arisen featuring the choice between QED PDF's with massless light quarks vs exact Feynman diagrams with short-distance quark masses and non-QED PDF's. The precision data should be able to settle these additional issues as it resolves the original discussion for the QCD case. Even higher precision may open further issues -- such is the nature of progress in precision applications of quantum field theory.
 \par
    We have pursued exact amplitude-based resummation realized on an event-by-event basis via shower/matrix element(ME) matched MC's in order to achieve enhanced precision for a given level of exactness while addressing the present paradigm in precision physics at the LHC and the futuristic FCC.  Currently, we have a realization of IR-improved parton showers in the Herwig6.5~\cite{hwg} environment in the MC Herwiri1.031~\cite{hwri} by two of us (BFLW and SAY). Via the MC@NLO~\cite{mcnlo} and the MG5\_aMC@NLO~\cite{mg5aMC} frameworks this MC  is elevated to the exact NLO shower/ME matched level as MC@NLO/Herwiri1.031~\cite{mcnlo-hwri} and MG5\_aMC@NLO/Herwiri1.031~\cite{mg5_amc-hwri}, respectively. We have the realization of IR-improved (IRI) Pythia8~\cite{cpc-py8} by one of us (BFLW) in the Pythia8~\cite{pythia8} environment,
with its corresponding NLO shower/ME matched  MG5\_aMC@NLO/IRI-Pythia8. More recently, two of us (BFLW and SAY), using the Herwig6.5 environment, have realized in the MC \KK{MC}-hh~\cite{kkmchh} exact ${\cal O}(\alpha^2L)$ CEEX EW corrections in a hadronic MC. \par
    As two of us (BFLW, SAY) have shown ~\cite{hwri,mcnlo-hwri}, IR-improvement in Herwig6.5 via Herwiri1.031 leads to improved precision in both the central $|\eta_\ell|\lesssim 2.5$ region for the ATLAS, CMS, D0 and CDF data and in the more forward region of LHCb where $2.0<\eta_\ell<4.5$. Here, $|\eta_\ell|$ is the lepton pseudorapidity in respective single $Z/\gamma^*$ production production with decay to lepton pairs. One of us (BFLW)~\cite{iri-semi-an} has shown the availability of the IR-improved semi-analytical paradigm for the latter processes. In what follows, we present an update the application of our methods in the analysis of LHC W+ n jets data, in the FCC discovery physics and in the interplay of IR-improved parton showers with exact ${\cal O}(\alpha^2L)$ CEEX EW corrections in \KK{MC}-hh.\par
    We note that 2017 was the 50th anniversary of the seminal paper by S. Weinberg~\cite{weinbg1} in which he formulated his foundational model of leptons in creating the spontaneously broken $SU_{2L}\times U_1$ EW theory~\cite{slg,aslm}, one of the key components of the SM, which we may now call the Standard Theory (ST) \footnote{See D.J. Gross, talk, {\it SM@50 Symposium}, Cleveland, OH, June, 2018.}. Progress on precision theory has been essential to the establishment of the ST\cite{weinbg1,slg,aslm,gw,pltzr}. As we celebrate 50 years of the ST (SM)~\cite{50yr-SM}, we are also obliged to look to the future with the FCC~\cite{fcc-study}, CLIC, ILC, or CEPC on the horizion~\cite{blondel-jnt}, where, for example, we note that the FCC 
\begin{figure}[h]
\begin{center}
\includegraphics[width=3.5in]{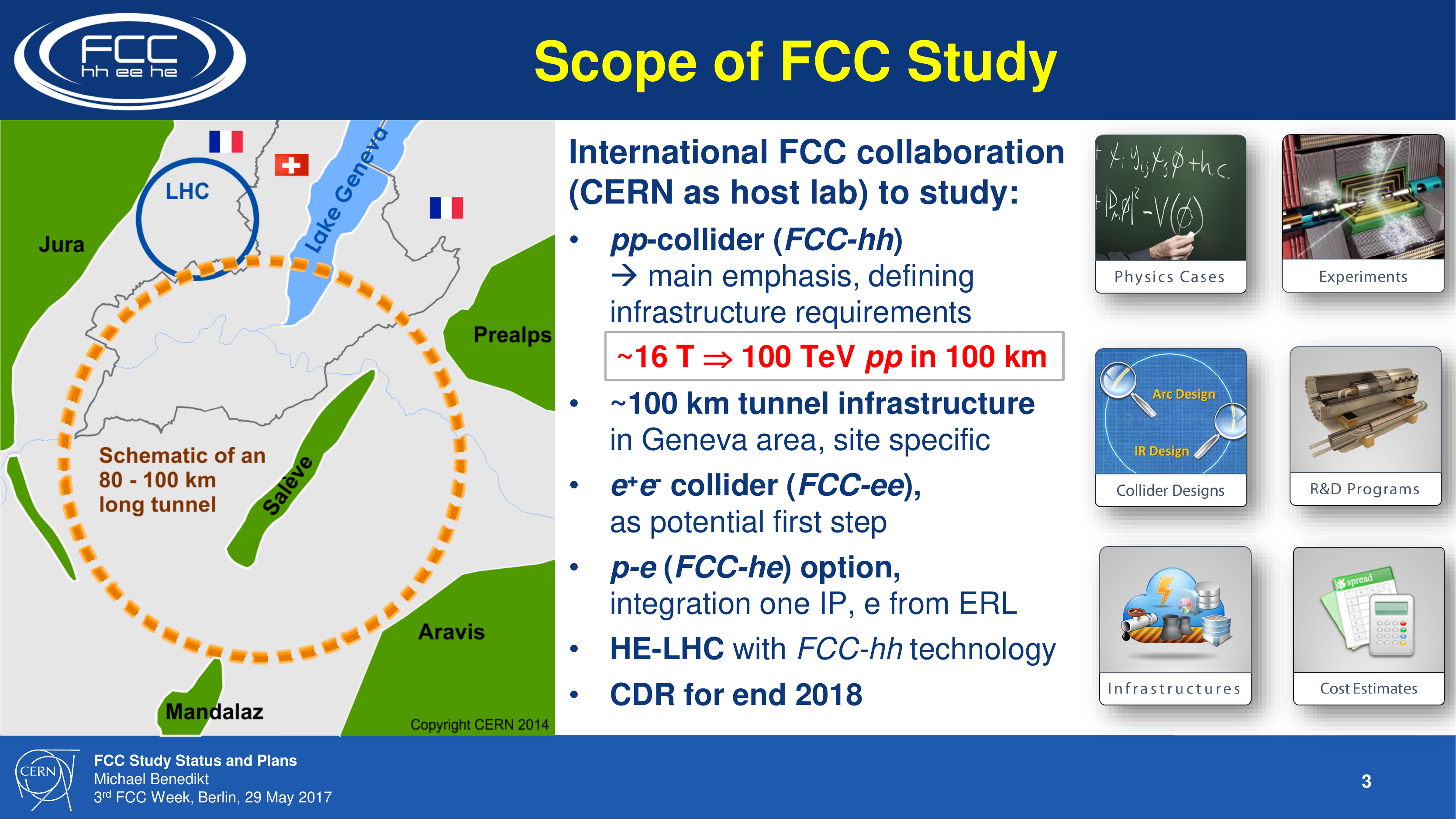}
\end{center}
\caption{\baselineskip=7pt FCC device planned for the future at CERN as depicted in Ref.~\cite{fcc-study}.}
\label{fig1}
\end{figure}
will feature a 100 TeV hadron collider and a tera-Z $e^+e^-$ colliding beam device, as illustrated in Fig.~\ref{fig1}. The success of the latter devices will be strongly correlated with  the progress of precision theory.\par
   The paper is organized as follows. We briefly review the parton shower implementation of exact amplitude-based resummation theory, after which we turn to the interplay of IR-improved DGLAP-CS QCD theory and shower/ME matched precision via comparisons with LHC data on $W + n$ jets and via predictions for FCC discovery physics. Finally, we discuss the interplay of IR-improved DGLAP-CS QCD theory and exact ${\cal O}(\alpha^2 L)$ CEEX EW corrections in single $Z/\gamma^*$ production at the LHC.\par  
   The parton shower implementation of exact amplitude-based resummation theory has as its starting point the master formula
\begin{eqnarray}
&d\bar\sigma_{\rm res} = e^{\rm SUM_{IR}(QCED)}
   \Large{\sum}_{{n,m}=0}^\infty\frac{1}{n!m!}\int\prod_{j_1=1}^n\frac{d^3k_{j_1}}{k_{j_1}} \cr
&\prod_{j_2=1}^m\frac{d^3{k'}_{j_2}}{{k'}_{j_2}}
\int\frac{d^4y}{(2\pi)^4}e^{iy\cdot(p_1+q_1-p_2-q_2-\sum k_{j_1}-\sum {k'}_{j_2})+
D_\rQCED} \cr
&{\tilde{\bar\beta}_{n,m}(k_1,\ldots,k_n;k'_1,\ldots,k'_m)}\frac{d^3p_2}{p_2^{\,0}}\frac{d^3q_2}{q_2^{\,0}},
\label{subp15b}
\end{eqnarray}
\small
where {\em new} (YFS-style) {\em non-Abelian} residuals 
{$\tilde{\bar\beta}_{n,m}(k_1,\ldots,k_n;k'_1,\ldots,k'_m)$} have {$n$} hard gluons and {$m$} hard photons. Definitions of the infrared functions ${\rm SUM_{IR}(QCED)}$ and ${ D_\rQCED}$ and of the residuals are given in Ref.~\cite{mcnlo-hwri}.  In the framework of shower/ME matching,  we have the replacements {$\tilde{\bar\beta}_{n,m}\rightarrow \hat{\tilde{\bar\beta}}_{n,m}$}. These replacements allow us, via the basic formula
\begin{equation}
{d\sigma} =\sum_{i,j}\int dx_1dx_2{F_i(x_1)F_j(x_2)} d\hat\sigma_{\rm res}(x_1x_2s),
\label{bscfrla}
\end{equation}
to proceed with connection to MC@NLO as explained in Ref.~\cite{mcnlo-hwri}.\par
\begin{figure}[h]
\begin{center}
\setlength{\unitlength}{0.1mm}
\begin{picture}(1600, 930)
\put( 390, 930){\makebox(0,0)[cb]{\bf (a)} }
\put(1140, 930){\makebox(0,0)[cb]{\bf (b)} }
\put(   00,500){\makebox(0,0)[lb]{\includegraphics[width=75mm]{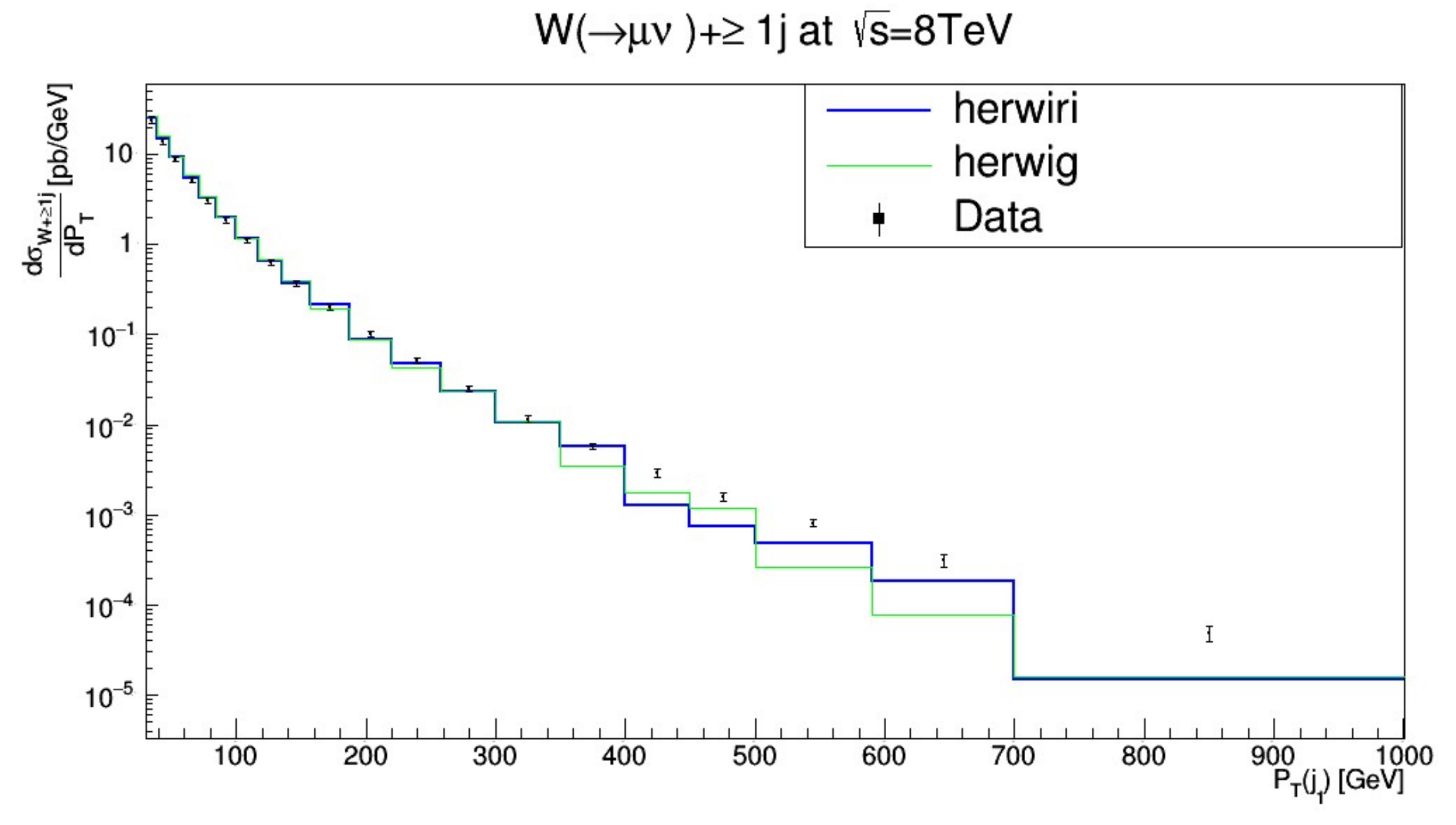}}}
\put( 750, 500){\makebox(0,0)[lb]{\includegraphics[width=75mm]{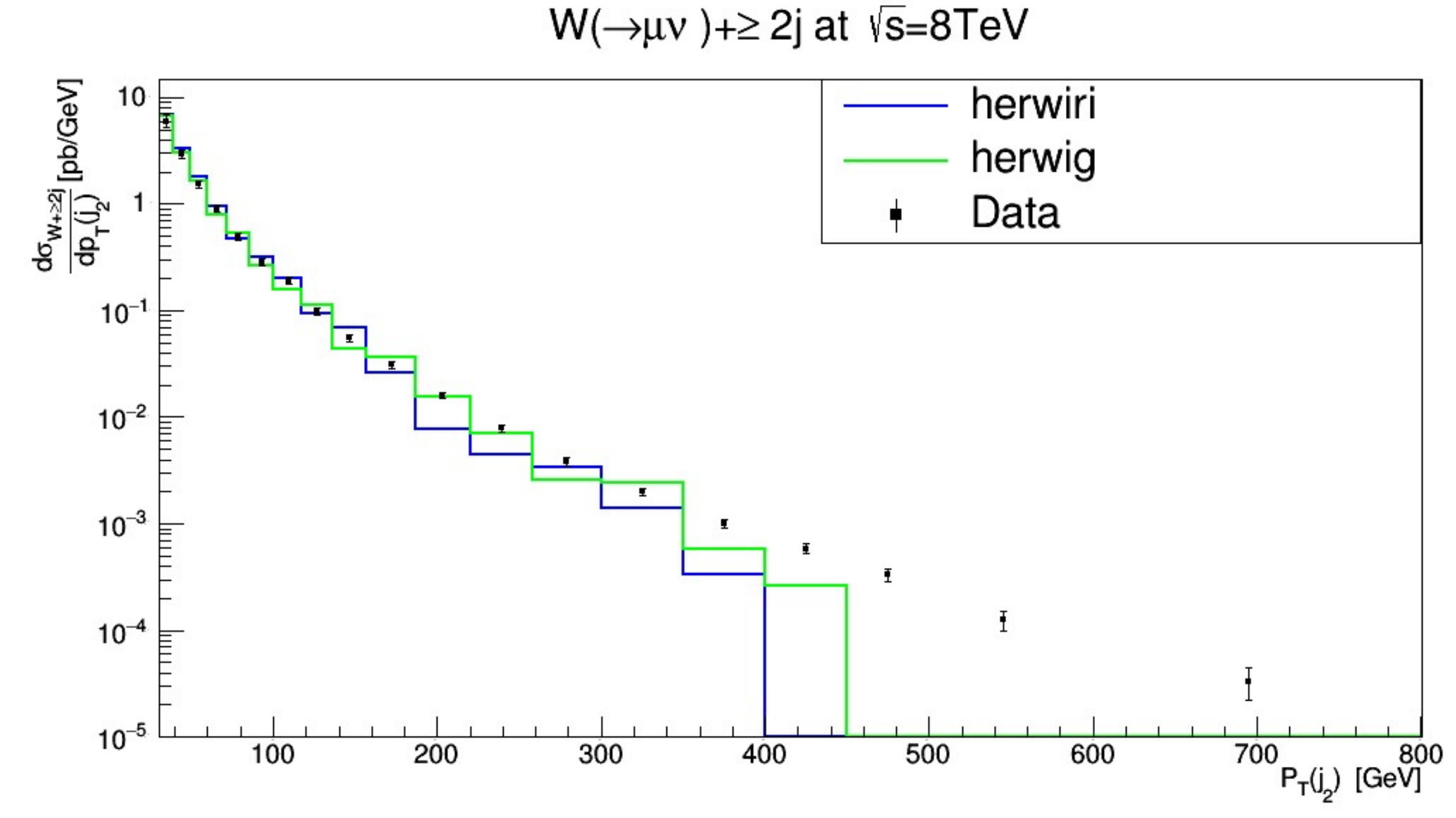}}}
\end{picture}
\end{center}
\vspace{-55mm}
\caption{\baselineskip=8pt Comparison of CMS 8TeV cms energy $W + \ge1$ jet (a) and $W + \ge2$ jets (b) data for the respective leading jet and second leading jet $p_T$ distributions and the IR-improved(herwiri)  and unimproved (herwig) exact NLO ME matched parton shower predictions.}
\label{figlhcj1-2}
\end{figure}
\begin{figure}[h]
\begin{center}
\setlength{\unitlength}{0.1mm}
\begin{picture}(1600, 930)
\put( 390, 930){\makebox(0,0)[cb]{\bf (a)} }
\put(1140, 930){\makebox(0,0)[cb]{\bf (b)} }
\put(   00,500){\makebox(0,0)[lb]{\includegraphics[width=75mm]{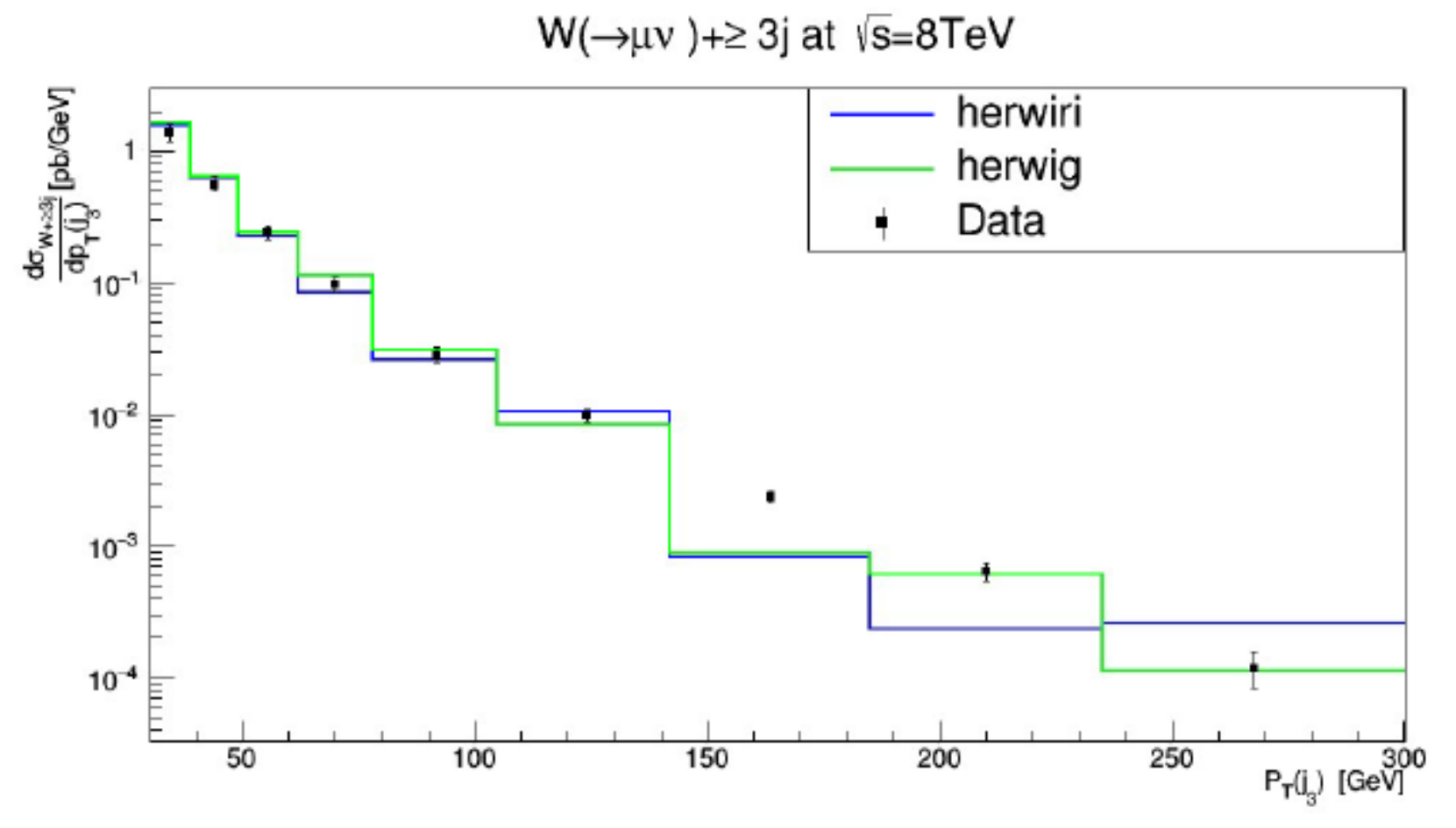}}}
\put( 750, 500){\makebox(0,0)[lb]{\includegraphics[width=75mm]{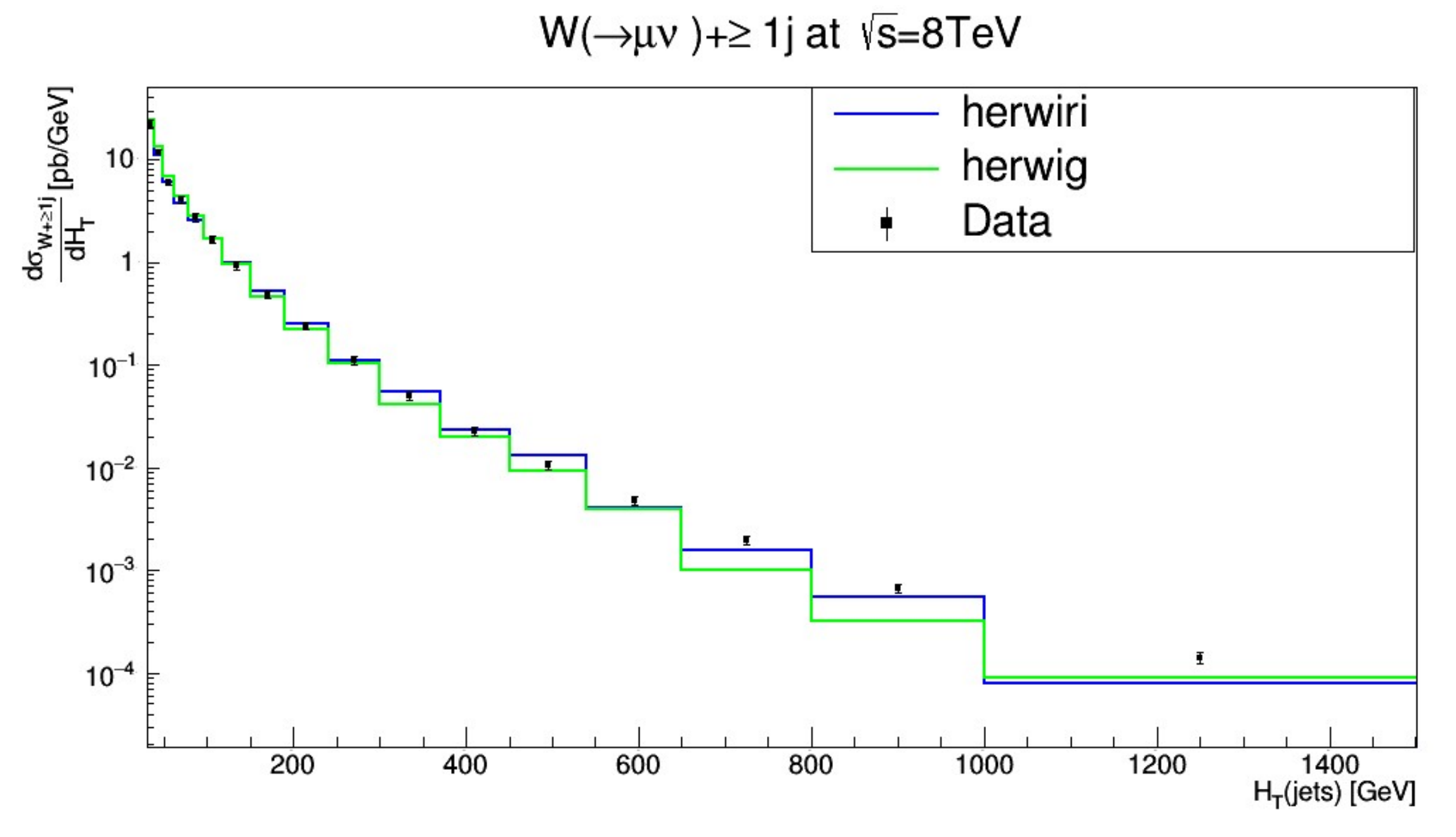}}}
\end{picture}
\end{center}
\vspace{-55mm}
\caption{\baselineskip=8pt Comparison of CMS 8TeV cms energy $W + \ge3$ jets (a) and $W + \ge1$ jet (b) data for the respective 3rd-leading jet  $p_T$ and  $H_T$ distributions and the IR-improved(herwiri)  and unimproved (herwig) exact NLO ME matched parton shower predictions.}
\label{figlhcj3ht}
\end{figure}

   Our (BS, BFLW) recent applications~\cite{bsh1,bsh2} make comparisons between the LHC data on $W+ n$ jets, n=1,2,3, and the exact NLO ME matched QCD parton shower predictions in the MG5\_aMC@NLO framework with the parton shower realized via Herwig6.5 and Herwiri1.031 respectively for the unimproved and IR-improved results. We illustrate such results in Figs.~\ref{figlhcj1-2} and ~\ref{figlhcj3ht} for the 
CMS~\cite{cms8tevdata} 8 TeV cms energy $W + \ge n$ jet data, n=1, 2, 3. Consistent with Refs.~\cite{mcnlo-hwri}, the IR-improved results are closer to the data for low $p_T(H_T)$.\par
   Two of us (BFLW and SAY) have investigated the effect of IR-improvement on the discovery reach of a standard candle process such as single $Z/\gamma^*$ production at the FCC using the predicted inclusive cross section for $Z/\gamma^*$ as a function of $p_{T,min}$ \footnote{This observable was suggested by M.L. Mangano, private communication, 2016.} which is explained and shown in Fig. 6 in Ref.~\cite{rdcr17}. In the latter figiure we plot predictions for the following: MG5\_aMC@NLO/A, A= Herwig6.5, Herwiri1.031, Herwig++ and Pythia8, all with the common renormalization and factorization scale of $M_Z$/2 and all with the common  renormalization and factorization scale of $H_T$/2 (denoted by 'UNFIX' in the legend in the figure) ; MG5\_aMC@NLO/Herwig6.5 and fixed order NLO both with the common renormalization and factorization $M_Z$; and, fixed order NLO with the common renormalization and factorization scale $H_T$/2. Here, $H_T$ is the sum of the transverse masses of the final state particles -- see Ref.~\cite{rdcr17} for the remaining details. Our results show that the dynamical scale choice makes a big difference in the expectations, the fixed-order NLO results agree, as it is expected, with the  MG5\_aMC@NLO/Herwig6.5 results for both of the  scale choices, and the IR-improved and unimproved predictions agree within the statistical uncertainties.\par  
   Finally, in Refs.~\cite{kkmchh2,kkmchh3,kkmchh4} two of us (BFLW and SAY), in collaboration with S. Jadach and Z. Was,  have analyzed the effects of the exact ${\cal O}(\alpha^2L)$ CEEX EW corrections in \KK{MC}-hh on the analysis of the $Z$ observables used in the ATLAS $M_W$ measurement in Ref.~\cite{atlasmw}. We find new effects that should be considered at or above the per mille level. For example, the new modulation we see in the lepton $p_T$ spectrum seems to match the trends in the data vs theory comparisons shown in Figs. 15 a and b in Ref.~\cite{atlasmw}. More specifically, we have found new effects sensitive to the transverse degrees of freedom in the ISR radiation as well as significant ISR effects in observables primarily sensitive to collinear ISR degrees of freedom, as we illustrate here in Table ~\ref{tab-1}.
\begin{table}[h]
\caption{\text{Illustrations of ISR and IFI Effects}}
\vbox{
{Numerical Results}
\begin{center}
\resizebox{\textwidth}{!}{%
\begin{tabular}{|l|c|c|c|c|c|c|}
\hline
 		& No ISR	& LuxQED ISR	& {\KK}MC-hh ISR & ISR$-$no ISR
		& With IFI & \%(IFI $-$ no IFI)\\
\hline
Uncut $\sigma$ 	& 939.86(1) pb	& 944.04(1) pb	& 944.99(2) pb	& 0.546(2)\%
		& 944.91(2) pb	& $-0.0089(4)$\%\\
Cut $\sigma$	& 439.10(1) pb	& 440.93(1) pb 	& 442.36(1) pb  & 0.742(3)\%
		& 422.33(1) pb	& $-0.0070(5)$\%\\
$A_{\rm FB}$	& 0.01125(2)	& 0.01145(2)	& 0.1129(2)	& $(3.9\pm2.8)
     \times 10^{-5}$	        & 0.01132(2)  	& $(2.9\pm 1.1)\times 10^{-5}$\\
$A_4$		& 0.06102(3)	& 0.06131(3)	& 0.06057(3)	&$(-4.4\pm0.5)
     \times 10^{-4}$            & 0.06102(3)	& $(4.5\pm 0.3)\times 10^{-4}$\\
\hline
\end{tabular}
}
\\[1em]
{The ISR and IFI effects on $A_{FB}$ is of order $10^{-5}$ while the effect on $A_4$ is of order $10^{-4}$\\ in \KK{MC}-hh. LuxQED gives an ISR effect on the order of $10^{-4}$ for both $A_{FB}$ and $A_4$.}
\end{center}
}
\label{tab-1}
\end{table}   
We are extending the analysis in Refs.~\cite{kkmchh2,kkmchh3,kkmchh3} to IR-improved showers with the new MC \KK{MC}-hh/Herwiri1.031 as explained and illustrated in Ref.~\cite{rdcr17}.\par
\section*{Acknowledgments} This work was partially supported by The Citadel Foundation. We thank the INP, Krakow, PL for computational resources. One of us (BFLW) thanks Prof. Gian Giudice for the support and kind hospitality of the CERN TH Department.
\setlength{\bibsep}{1.7pt}




\end{document}